# A NEW RADIATION-HYDRODYNAMICS CODE TO STUDY CORE-COLLAPSE SUPERNOVA EVENTS

M.L. Pumo[a,*] & L. Zampieri[b]

a) INAF - Osservatorio Astrofisico di Catania, Via S. Sofia 78, I-95123 Catania, Italy

b) INAF - Osservatorio Astronomico di Padova, Vicolo dell'Osservatorio 5, I-35122 Padova,

Italy

* Corresponding author, e-mail: mlpumo@oact.inaf.it

*Abstract*

With the aim of clarifying the nature of the core-collapse supernova events, we have developed a specifically tailored relativistic, radiation-hydrodynamics Lagrangian code, that enables us to simulate the evolution of the main observables (light curve, evolution of photospheric velocity and temperature) in these events.

The code features, some testcase simulations and the possible applications of the code are briefly discussed.

## INTRODUCTION

Core-collapse supernova (CC-SN) events are thought to be the final explosive evolutionary phase of stars having initial (i.e. at main sequence) mass larger than ~ 8-10 $M_\odot$ [e.g. 5, 7, 14]. Such as, the CC-SNe are fundamental probes of stellar evolution theories and can be used to verify the link among explosion mechanisms, nature of possible remnants, progenitors evolution, and environment around progenitors [e.g. 3, 4].

In addition to their intrinsic interest, CC-SNe are relevant to many astrophysical issues associated, for example, with the physical and chemical evolution of the galaxies, the nucleosynthesis processes of intermediate and trans-iron elements, the production of neutrinos, cosmic rays and gravitational waves [e.g. 1, 3, 6, 8, 12]. Moreover, CC-SNe seem to be particularly promising to measure cosmological distance, in addition to type Ia SNe [e.g. 17 and references therein].

Despite the importance of these explosive events in astrophysics, there are still basic questions to be answered, related to the extreme variety of CC-SNe displays (different energetics and amounts of ejected material, reflecting in heterogeneous light curves, spectra and evolution of photospheric velocity and temperature) and linked to the uncertainties in the modelling of stellar evolution and explosion mechanism [see e.g. 4, 9, 13 and references therein]. In particular the exact link between the physical properties of the explosion (ejected mass, explosion energy, stellar structure and composition at the explosion) and the observational characteristics is far from being well-established, and a "self-consistent" description of CC-SN events (from the quiescient evolutionary phases up the post-explosion evolution) is still missing.

In this context, the creation of specific modelling tools that link progenitor evolution and explosion models to the main observables (i.e. light curve, evolution of photospheric velocity and temperature) of CC-SNe is of primary importance for clarifying the nature of the CC-SN events. The developement of the relativistic, radiation-hydrodynamics code, described in this paper, represents a key step in this direction.

## CODE DESCRIPTION

The code, described in detail in [10] and [11], is a new and improved version of the one developed by [15], [16] and [2]. It solves the equations of relativistic, radiation hydrodynamics for a self-gravitating matter fluid interacting with radiation, taking into account the heating effects due to decays of radioactive isotopes synthesised in the CC-SN explosion. Moreover the code is capable of dealing with the fallback of material onto the compact remnant.

The main feature of this new version is the coupling in a fully implicit Lagrangian finite difference scheme of the gas energy equation and the equation for the radiation energy density with the radiative flux equation. This coupling allows for a major improvement in the numerical stability and overall computational efficiency of the code, especially during those evolutionary stages when fast motions of steep gradients occur (e.g. the radiative recombination phase).

As can be seen in Fig.s 1 and 2, the recombination front is very well resolved by the code.

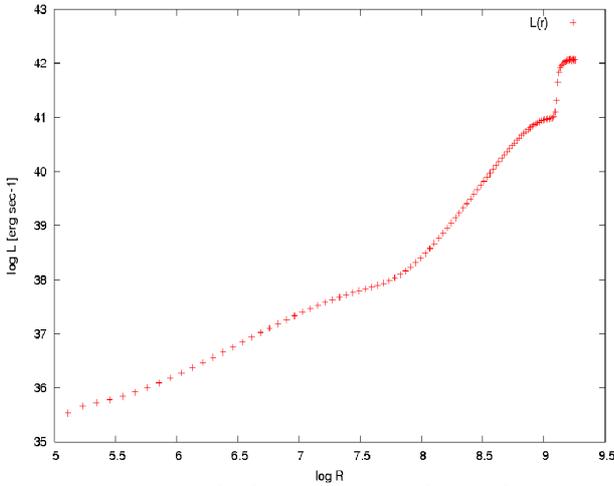

Figure 1: Radial profiles of the radiative luminosity as a function of radius (in units of the Schwarzschild radius for a compact object of 1.5$M_\odot$) at the "beginning" of the recombination phase (less than 5% by mass of the envelope has recombined) for a model having total mass of 16 $M_\odot$, initial radius of $3\times10^{12}$ cm, total (kinetic plus thermal) energy of 1 FOE (equal to $10^{51}$ erg), and $^{56}$Ni mass of 0.070 $M_\odot$. A logarithmic scale is used. The sharp boundary, where also the photosphere is located, marks the position of the recombination front (at log R ~ 9.15).

The numerical stability makes it possible to follow the post-explosion evolution of a CC-SN event in its "entirety", from the breakout of the shock wave at the stellar surface up to the so-called nebular stage, in which the ejected envelope has recombined and the energy budget is dominated by the radioactive decays of the heavy elements synthesised in the explosion.

Consequently, the code is able to simulate the post-explosion evolution of the main observables in CC-SN events for a given set of initial parameters (kinetic and thermal energy, envelope mass, radius, amount and distribution of $^{56}$Ni) that are linked to the physical conditions of the CC-SN progenitor at explosion.

Fig. 3 shows an example of the output of the code that can be directly compared with the observations through a best fitting procedure that is based on chi-square minimization and provides in output the physical parameters of the CC-SN progenitor at explosion [see 17 for details].

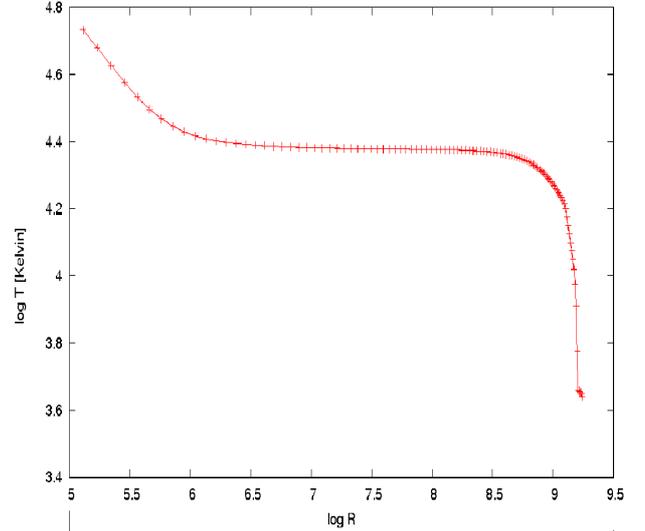

Figure 2: Same as Fig. 1, but for the radial profiles of the temperature.

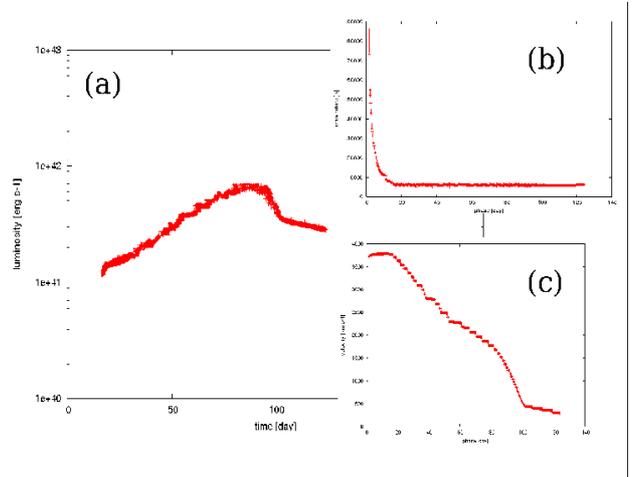

Figure 3: Light curve (panel a), evolution of the photospheric velocity (panel b) and temperature (panel c) for a model having total mass of 16$M_\odot$, initial radius of $3\times10^{12}$ cm, total energy of 1 FOE, and $^{56}$Ni mass of 0.070 $M_\odot$.

## WORK IN PROGRESS AND FUTURE DEVELOPMENTS

We are now working at computing fine grids of models evolved from post-explosion configurations. We intend to carry out detailed investigations of the evolution of CC-SN events, in order to determine how their main observables depend on the physical conditions of the CC-SN progenitor at the explosion.

Fig.s 4 and 5 show the preliminary results of some of these calculations.

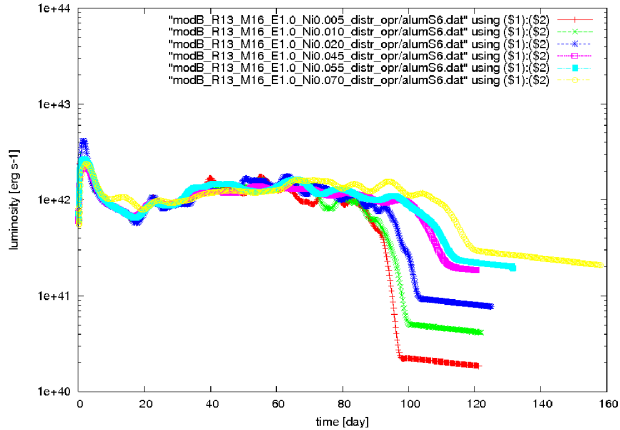

Figure 4: Effects of varying the $^{56}$Ni mass on the light curve for models having the same total mass (16 $M_\odot$), initial radius ($3 \times 10^{13}$ cm) and total energy (1 FOE). The tracks correspond to $^{56}$Ni masses equal to 0.005 $M_\odot$ (red line), 0.010 $M_\odot$ (green line), 0.020 $M_\odot$ (blue line), 0.045 $M_\odot$ (purple line), 0.055 $M_\odot$ (cyan line), and 0.070 $M_\odot$ (yellow line).

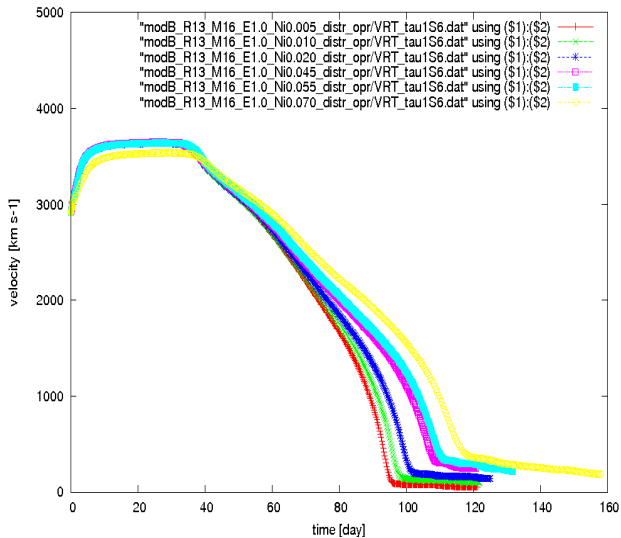

Figure 5: Same as Fig. 4, but for the evolution of the photospheric velocity.

Moreover we are starting to interface our new relativistic, radiation-hydrodynamics code with stellar evolution, nucleosynthesis and spectral synthesis codes, in order to develop a sort of "CC-SNe Laboratory" which will allow us to describe in a self-consistent way the evolution of a CC-SN event from the evolutionary stages preceeding main sequence up to the post-explosive phases as a function of initial mass, metallicity and mass loss history of the CC-SN progenitor.

*Acknowledges*

M.L.P. thanks the organizers of the *VII Incontro dei Gruppi Italiani di Astrofisica Nucleare Teorica e Sperimentale (GIANTS2010)* for the very fruitful meeting.

We acknowledge the *TriGrid VL* project and the *INAF-Astronomical Observatory of Padova* for providing computer facilities.